\documentstyle[prl,twocolumn,aps]{revtex}

\begin{document}

\draft
\twocolumn[\hsize\textwidth\columnwidth\hsize\csname
@twocolumnfalse\endcsname]

\noindent
{\large {\bf Comment on ``Correlation of Tunneling Spectra in 
Bi$_2$Sr$_2$CaCu$_2$O$_{8+ \delta}$ with the Resonance Spin Excitation''}}

\vspace{3mm}
In a recent Letter Zasadzinski et al. [1] reported scanning tunneling 
spectroscopy on Bi$_2$Sr$_2$CaCu$_2$O$_{8+ \delta}$ (Bi2212) and 
an interpretation of the data. In this Letter, the tunneling data are practically 
identical to their previously published data [2] and, therefore, do not present 
a new contribution to the existing literature. The main point of their latter 
work was to present a new interpretation of the data, namely, to link 
dips in tunneling conductances with the so-called magnetic resonance peak 
observed in inelastic neutron scattering (INS) measurements [3]. 

The tunneling conductances obtained at low temperature in Bi2212 have a 
well-defined structure---the presence of quasiparticle peaks, dips and humps. 
In a superconductor-insulator-normal metal (SIN) junction, the conductance 
peaks are located at a bias $V_{peak} = \Delta /e$, while in a 
superconductor-insulator-superconductor (SIS) junction, they appear at 
$V_{peak} = 2 \Delta /e$, where $\Delta$ is the energy gap, and $e$ is the 
electron charge. 
The authors of Ref. [1] measured the difference $\Omega = e(V_{dip} -  
V_{peak})$ in SIS conductances as a function of doping level $p$, where 
$V_{dip}$ is the dip bias. They found that $\Omega (p) \simeq E_r(p)$, where 
$E_r$ is the energy at which the magnetic resonance mode is situated in INS 
spectra. By analogy with phonon structures in tunneling conductances obtained 
in conventional superconductors, they concluded that the dips in tunneling 
conductances of Bi2212 are caused by the magnetic excitation seen by INS. 
Below, we show that their interpretation of the data is incorrect. 

First, earlier [4] and recently [5] it was {\em experimentally} shown 
that the dips in tunneling conductances ``have no physical meaning'' and 
``appear naturally due to a superposition of two contributions'' (peaks and 
humps). Angle-resolved photoemission (ARPES) measurements performed 
in Bi2212 fully support the latter scenario [6]. The correlation 
$\Omega (p) \simeq E_r (p)$ found in Ref. [1] for Bi2212 is a chance 
coincidence. 

Secondly, the authors of Ref. [1] wrote: ``It should be noted that a similar dip 
has been observed in the superconducting tunneling spectra of a heavy 
fermion superconductor [7] which has also been linked to a peak that develops 
in the spin excitation spectrum. Thus, a spin fluctuation mechanism may have 
a more general relevance to superconductors beyond the high $T_c$ cuprates.'' 
Indeed, SIN tunneling conductances obtained in the heavy fermion 
UPd$_2$Al$_3$ [7] are reminiscent of those measured in Bi2212, and it is 
generally believed that spin fluctuations mediate superconductivity in 
UPd$_2$Al$_3$. Then, if the 
correlation $\Omega \simeq E_r$ is valid for Bi2212, it must be valid for 
UPd$_2$Al$_3$ too. However, this is not the case. Experimentally, in 
UPd$_2$Al$_3$ $\Omega = eV_{dip} - eV_{peak} \simeq$ 0.88 - 0.235 = 0.645 
meV [7], while $E_r$ = 1.5--1.65 meV  [8--10]. Thus, in UPd$_2$Al$_3$ the 
value $\Omega$ = 0.645 meV is more than twice smaller than the value 
$E_r$ = 1.5--1.65 meV. Therefore, the dips in tunneling conductances 
obtained in UPd$_2$Al$_3$ can not be caused by the spin excitation 
associated with the magnetic resonance mode in INS spectra. 

Thirdly, return to cuprates: the authors of Ref. [1] wrote: ``As a final 
comment, we note that similar dip feature have been observed in the 
tunneling spectra of Tl$_2$Ba$_2$CuO$_6$ indicating that the neutron 
resonance ought to be observed in a cuprate with a single Cu-O layer per 
unit cell.'' Experimentally, in near optimally-doped Tl$_2$Ba$_2$CuO$_6$ 
$\Omega \simeq$ 16 meV [11], while $E_r \simeq$ 47 meV [12]. Thus, in 
Tl$_2$Ba$_2$CuO$_6$ the value $\Omega \simeq$ 16 meV is three times 
smaller than the value $E_r \simeq$ 47 meV. Therefore, the dips in 
tunneling conductances obtained in Tl$_2$Ba$_2$CuO$_6$ can not be caused 
by the spin excitation. 

To summarize, undoubtedly, the spin excitation manifesting itself 
as a resonance peak in INS spectra is an important part of the mechanism 
of unconventional superconductivity in cuprates and heavy fermions [5]. 
However, the dips in tunneling conductances obtained in these unconventional 
superconductors, as shown above, have nothing to do with this magnetic 
excitation. 

\vspace{5mm} 
\noindent
A. Mourachkine \\
\hspace*{2.5mm} Universit\'e Libre de Bruxelles, CP-232, \\
\hspace*{2.5mm} Blvd. du Triomphe, B-1050 Brussels, Belgium

\vspace{3mm}

\end{document}